\documentclass[5p]{elsarticle}

\usepackage{amsmath}
\usepackage{amsthm}
\usepackage{graphicx}
\usepackage{epsfig}

\usepackage[french]{babel}
\usepackage{subfigure}
\usepackage{stmaryrd}

\def\div{{\nabla\cdot}}
\def\rot{{\nabla\times}}
\def\d{{\rm d}}
\journal{ArXiv}

\begin{document}

\begin{frontmatter}

\title{A second-order pressure-accurate finite-difference scheme for the Stokes problem with rigid non-conforming boundaries}

\author[LN]{Abdelkader Hammouti}
\ead{abdelkader.hammouti@lcpc.fr}
\author[LN]{Ana\"el Lema\^{\i}tre}
\ead{anael.lemaitre@ifsttar.fr}

\address[LN]{Universit\'e Paris Est -- Laboratoire Navier\\
ENPC-ParisTech, LCPC, CNRS UMR 8205\\
2 all\'ee Kepler, 77420 Champs-sur-Marne, France}

\begin{abstract}
We present a finite-difference scheme which solves the Stokes problem in the presence of curvilinear non-conforming interfaces and provides second-order accuracy on physical field (velocity, vorticity) and especially on pressure. The gist of our method is to rely on the Helmholtz decomposition of the Stokes equation: the pressure problem is then written in an integral form devoid of the spurious sources known to be the cause of numerical boundary layer error in most implementations, leading to a discretization which guarantees a strict enforcement of mass conservation. The ghost method is furthermore used to implement the boundary values of pressure and vorticity near curved interfaces.
\end{abstract}


\end{frontmatter}

\section{Introduction}

Obtaining second-order accuracy on pressure, in numerical simulations of viscous incompressible flows in the presence of rigid interfaces, remains a largely open problem, as standard algorithms introduce a numerical boundary layer that pollutes the pressure field~\cite{ELiu1995,Dagan2003}. These errors, which arise in the simplest cases -- near flat walls conforming with the underlying mesh in finite element methods~\cite{GuermondMinevShen2006} -- become daunting when interfaces cannot be aligned with the computational grid. This latter issue, however, is critical to the development of Cartesian grid methods -- a quite active field -- which seek increased efficiency and flexibility by avoiding any remeshing even when dealing with curved boundaries.

The origin of pressure errors in direct numerical simulations can be examined by simply considering the incompressible Stokes problem:
\begin{subequations}\label{eqn:Stokes}
\begin{align}
\rho \, \mathbf{u}_{t}  &=  -\nabla p + \mu\Delta \mathbf{u} + \mathbf{f}\label{eqStokes0}\\
\div\mathbf{u} &= 0\label{divu=0}\\
\mathbf{u} &= \mathbf{u}_b \quad\text{on}\quad\partial\Omega
\end{align}
\end{subequations}
on a domain $\Omega$, with $\rho$ and $\mu$ the fluid density and viscosity, $\mathbf{u}$ and $p$, velocity and pressure, $\mathbf{f}$ bulk forces, and $\mathbf{u}_b$ the velocity boundary condition (BC). (We will use throughout the standard indicial notation for partial derivatives). The key problem is that pressure has no explicit expression as a function of others fields, but is determined implicitly within the Stokes problem: it is a Lagrange multiplier associated with the condition of incompressibility.

The strategies developed to cope with this difficulty can be sorted into two main families. (I) Either mass conservation is not strictly enforced -- as in fractional step or projection methods~\cite{Temam1968,Chorin1968} -- but attempt is made to reconstruct the physical pressure from fields computed at intermediate times~\cite{GuermondMinevShen2006,kimmoin1985,Brown2001,Bell1985,QuartapelleNapolitano1986}. (II) Or, the continuity condition is strictly enforced via e.g. appropriate polynomial formulations~\cite{Botella1996}, or by relying on potential~\cite{Ben-Artzi2006,Benrichou2005}, vorticity~\cite{ELiuV1995, Ben-Artzi2001}, or mixed potential-vorticity~\cite{Calhoun2002,Russell2003} formulations. The interest of type (I) methods is that they are usually based on the usual ``velocity-pressure'' formulation of the Stokes problem~\cite{ELiu1995,ELiu2001}, which involves only second order derivatives; however, they provide $O(\Delta t^{1/2})$~\cite{Temam1968,Chorin1968}, $O(\Delta t)$~\cite{Bell1985,VanKan1986}, or at best $O(\Delta t^{3/2})$~\cite{GuermondMinevShen2006,kimmoin1985,Guermond1999,KarniadakisIsraeliOrszag1991} convergence for $p$. Type (II) strategies provide second order accuracy on pressure,~\cite{Ben-Artzi2006,Calhoun2002} but rely either on matrix formulations of higher rank (the bi-Laplacian of potential formulations) or on complex, integral forms of the boundary conditions (in pure vorticity methods), which can become quite untractable with complicated boundary geometries.

Here, we present a method which both (i) is based on the natural velocity and pressure fields, and (ii) guarantees a strict enforcement of mass conservation. This is performed by constructing the discrete problem -- in finite differences -- on the basis of a Helmholtz decomposition of the Stokes problem (detailed in Sec.~\ref{sec:helmholtz}), leading quite naturally to a mixed velocity-pressure-potential-vorticity formulation. Our solver will be shown to achieve second order accuracy on all the physical fields including pressure even near curved non-conforming interfaces.

As in all numerical schemes of type (I), the critical step in our algorithm will be to solve for $p$. This usually is done via the following equation:
\begin{equation}
\label{eqPressure}
\Delta p = \div\mathbf{f}
\end{equation}
which comes after taking the divergence of~(\ref{eqStokes0}) and using $\div\Delta \mathbf{u} = \Delta\div\mathbf{u} = 0$. This expression, however, brings up several problems.

A first difficulty arises because it is tough to ensure that the discrete operators $\div^{(d)}$, $\mathbf{\nabla}^{(d)}$, $\rot^{(d)}$, and $\Delta^{(d)}$ verify the correct commutation relations: in naive implementations, $\div^{(d)}\Delta^{(d)}\mathbf{u}$ is non-vanishing near boundaries, which amounts to introducing spurious sources in equation~(\ref{eqPressure}). This problem is dealt with, on square grids, by using the rotational form of the Laplacian~\cite{Guermond2001}, but we should anticipate further intricacies when dealing with non-conforming boundaries.

A second, more serious, difficulty arises from the fact that, even though pressure verifies the Poisson equation~(\ref{eqPressure}), it cannot be simply seen as the solution of a plain Poisson (Dirichlet or Neumann) problem~\cite{GuermondMinevShen2006,Guermond1999}. The boundary condition on pressure indeed has no explicit form. The Stokes equation itself does introduce constraints on the vector components of $\nabla p$ on $\partial\Omega$, yet viewing them as boundary conditions for the pressure problem raise tremendous difficulties:
\begin{itemize}
\item setting $\mathbf{n}\cdot\nabla p$, with $\mathbf{n}$ the normal vector to an interface, is a Neumann condition; it is associated with mass flux through the interface i.e. with incompressibility
\item setting $\mathbf{t}\cdot\nabla p$, for any vector $\mathbf{t}$ tangent to the interface, fixes the pressure boundary values up to a constant, i.e. sets a Dirichlet condition, and is associated with the no slip boundary condition on velocity.
\end{itemize}
The Stokes equation thus sets both a Dirichlet and a Neumann boundary conditions on the pressure problem. It is a property of the Stokes problem that these two conditions are compatible, i.e. lead to the same solution for pressure. Most pressure-based spatial discretizations, however, introduce inconsistencies: solving the Poisson equation~(\ref{eqPressure}) with either boundary condition lead to different solutions as originally observed by Gresho {\it et al}~\cite{GreshoSani1987}. Guaranteeing that both conditions are simultaneously enforced thus remains a difficult issue and a recurrent theme in the literature hinges around the appropriate choice of either one~\cite{GuermondMinevShen2006,ELiu1996}: in most cases, the no-slip condition and the constraints on mass transport through the interface are not simultaneously enforced, which introduces large, uncontrolled, errors and prevents proper convergence of numerical approximations for pressure~\cite{ELiu1995}.

We will show that by performing a Helmholtz decomposition of the Stokes equation it is possible to formulate the pressure problem in an integral form which guarantees that its boundary conditions are well-posed and consistent with mass conservation especially near boundaries. To cope with curved boundaries within a finite difference scheme, we will rely on the ghost fluid method~\cite{FedkiwAslamMerrimanOsher1999}, which has been developed to take into account situations where the solutions to an elliptic problem and their derivatives have jumps at sharp sub-grid interfaces. In this method, interface conditions are implemented through the discretized matrix problem itself, in contrast with e.g. immersed boundary methods where they are represented via a set of localized sources. The ghost fluid method was first applied to two-phase incompressible flows~\cite{FedkiwAslamMerrimanOsher1999}, then to Poisson problems with Dirichlet or mixed boundary conditions~\cite{LiuFedkiwKang2000,GibouFedkiwCheng2002} on fixed rigid interfaces.

In Section~\ref{sec:pressure} we construct step by step the discrete pressure problem and analyze its convergence properties. Several outstanding questions will then remain regarding the incorporation of this solver into the Stokes problem: the computation of the boundary conditions for $p$ from the instantaneous velocity field; the formulation of a consistent rotational form of the Laplacian. They are discussed in Section~\ref{sec:implementationStokesPb} and shown to permit the construction of a scheme with second-order accuracy on all fields, including pressure. Test cases are finally presented in Section~\ref{sec:testcases}.

\section{Solving for pressure}
\label{sec:pressure}

\subsection{Helmholtz form of the Stokes problem}
\label{sec:helmholtz}

\subsubsection{Mass conservation \& Helmholtz decomposition}
\label{sec:helmholtz3d}

Our approach is based on the strict enforcement of mass conservation, which is achieve by devising the discretized problem around the Helmholtz structure of the Stokes equation. It is inspired by the effectiveness of using the rotational form for the Laplacian~\cite{GuermondMinevShen2006,KarniadakisIsraeliOrszag1991,IsraeliOrszagDeville1986}: writing $\mu\Delta\mathbf{u}=-\mu\rot\mathbf{\omega}$ is innocent from the viewpoint of continuum equation, yet it permits to avoid introducing spurious sources in the discretization of the Stokes equation~(\ref{eqStokes0}). The reason is that the constraint $\div^{(d)}\Delta^{(d)}\mathbf{u}=0$ is difficult to enforce numerically (it is non-local, i.e. involves an extended stencil), while guaranteeing $\div^{(d)}\rot^{(d)}\equiv0$ is relatively easy with proper definitions of the curl and divergence operators.

In the same spirit, we seek to write the terms $-\nabla p + \mathbf{f}$ in a rotational form, to insure that the computation of the pressure field respects incompressibility by construction. To do so, let us first recall that in both 2 and 3 dimension, the velocity field -- being solenoidal -- can be written as $\mathbf{u}=\rot\mathbf{A}$, with $\mathbf{A}$ a potential vector (we do not fix the gauge as yet). Introducing the vorticity field $\mathbf{\omega}=\rot\mathbf{u}$, the Stokes equation can therefore be written:
\begin{equation}
\label{eq:h2}
\rho\,\rot\mathbf{A}_t+\mu\,\rot\mathbf{\omega} + \nabla p = \mathbf{f}
\end{equation}
which can indeed be seen as a Helmholtz decomposition for the field of bulk forces. Equation~(\ref{eq:h2}) is the basis of our representation of the pressure problem.

\subsubsection{The 2D case}

As the general treatment of the pressure problem in 3D is somewhat involved, we focus most of this work on the 2D case: the discussion of Section~\ref{sec:conclusion:3D} will show that the analysis of the Helmholtz form of the Stokes problem and the methods developed here directly extrapolate to 3D problems.

The 2D velocity field is denoted $\mathbf{u}=(u,v)$. As usual, in two dimensions, the vorticity $\omega\mathbf{e}_z$ and potential $\mathbf{A}=\psi\mathbf{e}_z$ are effectively scalar quantities. The curl operator takes two different forms when it is applied on a 2D vector field or on a scalar potential: we thus write $\omega=\rot\mathbf{u}=v_x-u_y$, but $\mathbf{u}=(\psi_y,-\psi_x)=\nabla^\perp\psi$.

To better expose the Helmholtz structure of the Stokes problem
\begin{equation}
\label{eq:stokes:2d}
\rho\,\nabla^\perp\psi_t=-\nabla p-\mu\nabla^\perp\omega+\mathbf{f}
\end{equation}
we introduce the field
\begin{equation}
\phi=\rho\,\psi_t+\mu\omega
\end{equation}
so as to write
\begin{equation}
\label{eq:h:2d}
\nabla^\perp \phi+\nabla p = \mathbf{f}
\end{equation}
This equation reveals some form of ``conjugation'' between $p$ and $\phi$ that is further evidenced by constructing integral equations relating these two fields. Let us, for this purpose, consider a curve $\Gamma\subset\Omega$, running from point $A$ to $B$; with $\mathbf{t}=(t_x,t_y)$ the normalized tangent vector; integrating along $\Gamma$ yields:
\begin{equation}\label{eqn:integrale:p}
p(B)-p(A)= -\int_{\Gamma}\mathbf{t}\cdot\nabla^\perp\mathbf{\phi}+\int_{\Gamma} \mathbf{f}\cdot\mathbf{t}
\end{equation}
To check that this relation defines $p$ as a univalued function, we next need to ascertain that in the case $A=B$, i.e. when $\Gamma$ is a closed loop, the rhs of this equation vanishes. Taking $\Gamma$ to loop counterclockwise, denoting $\mathbf{n}=(-t_y,t_x)$ the outer normal vector, using the Stokes theorem and $\mathbf{t}\cdot\nabla^\perp\phi=\mathbf{n}\cdot\nabla\phi$, we find:
\begin{equation}\label{eqn:integrale:phi}
-\oint_{\Gamma}\mathbf{t}\cdot\nabla^\perp\mathbf{\phi}+\oint_{\Gamma} \mathbf{f}\cdot\mathbf{t}=\int_S \left(-\Delta\phi-\rot\mathbf{f}\right) \equiv 0
\end{equation}
where $S$ in the surface enclosed by $\Gamma$. The last integral vanishes because as found after taking the curl of~(\ref{eq:h:2d}): $\Delta \phi=-\rot\mathbf{f}$.

In perfect analogy, path integrals of the form
\begin{equation}
\label{integral:p:2d}
\phi_B-\phi_A=\int_\Gamma \mathbf{n}\cdot\left(-\nabla p+\mathbf{f}\right)\,\d s
\end{equation}
define $\phi$ -- up to an irrelevant constant -- from $p$, as in particular:
\begin{equation}
\label{integral:p:2d:closed}
0=\oint_\Gamma \mathbf{n}\cdot\left(-\nabla p+\mathbf{f}\right)\,\d s
\end{equation}
when $\Gamma$ in a closed path.

These latter integral equations,~(\ref{integral:p:2d}) and~(\ref{integral:p:2d:closed}), are the basis of our implementation. By posing the pressure problem as a set of path integrals of this form, without actually solving for $\phi$, we guarantee that $-\nabla p + \mathbf{f}$ is the curl of an unknown field ($\phi$); this in turn guarantees that $\div(-\nabla p + \mathbf{f})$ vanishes strictly in the discretized the problem. Furthermore, this integral pressure problem is fully determined by the boundary values of $\phi$. The solution $p$ of this integral problem thus automatically verifies the dual Neumann and Dirichlet boundary conditions set by the Stokes equation~(\ref{eq:stokes:2d}): the agonizing choice~\cite{ELiu1995} of one versus another is thus resolved.

In the rest of this section, we assume that the boundary values of $\phi$ are known, and show how~(\ref{integral:p:2d}) and~(\ref{integral:p:2d:closed}) can then be discretized using finite differences into a well-posed numerical problem for $p$.

\subsection{The MAC grid}

\begin{figure}[!ht]
\begin{center}
\includegraphics{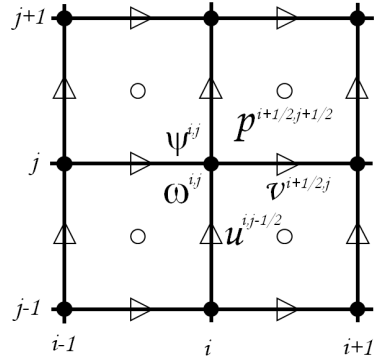}
\caption{The MAC grid} \label{fig:MAC}
\end{center}
\end{figure}
Our space discretization uses the MAC scheme, which involves staggered grids as displayed on Fig.~\ref{fig:MAC}, with cell size $\Delta x\times\Delta y$. We denote $h\propto\Delta x\propto\Delta y$ a characteristic discretization scale. The pressure variables are positioned at the center of the cells (``$\circ$'' points); the first and second components of velocity, $u$ and $v$, on the edges (``$\triangle$'' and ``$\triangleright$'', resp.); and the fields $\psi$, $\omega$, and $\phi=\rho\,\psi_t+\mu\,\omega$ at the corners of the cells. These choices guarantee that on regular grid points all computations of derivatives only require centered differences.

Integer indices $i$ and $j$ tag the location of the grid lines in the $x$ and $y$ directions, respectively: pairs of integer indices thus mark the location of the $\psi$ and $\omega$-type fields. Edges are numbered by the index of their midpoint, of the form $i+\frac{1}{2},j$ and $i,j+\frac{1}{2}$ for horizontal and vertical edges, respectively.

We introduce the differential operator $\delta_x,\delta_y$ defined by their action of discrete fields $a^{\alpha,\beta}$ -- with here $\alpha,\beta$ integers or half-integers:
\begin{align*}
\left(\delta_x a\right)^{\alpha,\beta}&=\frac{a^{\alpha+\frac{1}{2},\beta}-a^{\alpha-\frac{1}{2},\beta}}{\Delta x}\\
\left(\delta_y a\right)^{\alpha,\beta}&=\frac{a^{\alpha,\beta+\frac{1}{2}}-a^{\alpha,\beta-\frac{1}{2}}}{\Delta y}
\end{align*}
The discrete gradient operator (which normally applies of $p$-type fields) is thus $\nabla^{(d)} = (\delta_x,\delta_y)$, the discrete divergence (which applies on $(u,v)$-type fields) has the form $\div^{(d)}\mathbf{u}=\delta_x u+\delta_y v$, and the discrete rotational operator (which applies of $\phi$-type fields) is $\nabla^{\perp (d)}=(\delta_y,-\delta_x)$.
We finally denote $\Delta^{(d)} = \delta_x^2+\delta_y^2$ the discrete Laplacian.
With these definitions, it is clear that
\begin{equation}
\label{eq:divrot}
\div^{(d)}\nabla^{\perp (d)}\equiv0
\end{equation}
at regular grid points.

\subsection{Paths, edges, and segments}

Our discretization of the pressure problem is constructed by writing integrals of the form~(\ref{integral:p:2d}) or~(\ref{integral:p:2d:closed}) on all possible paths $\subset\Omega$ that can be formed using the edges of MAC grid cells. These paths can always be decomposed as series of rectilinear segments. We define accordingly:
\begin{itemize}
\item a \emph{regular segment}: any edge of the MAC grid that entirely belongs to the fluid domain $\Omega$
\item an \emph{irregular segment}: the part $\subset\Omega$ of an irregular cell edge, i.e. an edge intersected by the domain boundary
\end{itemize}
Given an edge indexed $\alpha,\beta$, $\theta^{\alpha,\beta}\in[0,1]$ denotes the fraction of it that lies within the fluid domain and $\gamma^{\alpha,\beta}$, the segment it supports -- i.e. its intersection with $\Omega$. The segments' lengths are thus:
\begin{align*}
|\gamma^{i\pm1/2,j}|&=\theta^{i\pm1/2,j}\Delta x\\
|\gamma^{i,j\pm1/2}|&=\theta^{i,j\pm1/2}\Delta y
\end{align*}
By definition, regular edges ($\theta=1$) have both ends lying at the corner of a cell edge; irregular edges ($0<\theta<1$) have one such regular end point and an irregular one, which lies right at the interface~\footnote{We disregard the rare cases when a cell edge would be crossed several times by the boundary as these become irrelevant for a sufficiently small discretization step}.

\begin{figure}[!ht]
\begin{center}
\includegraphics{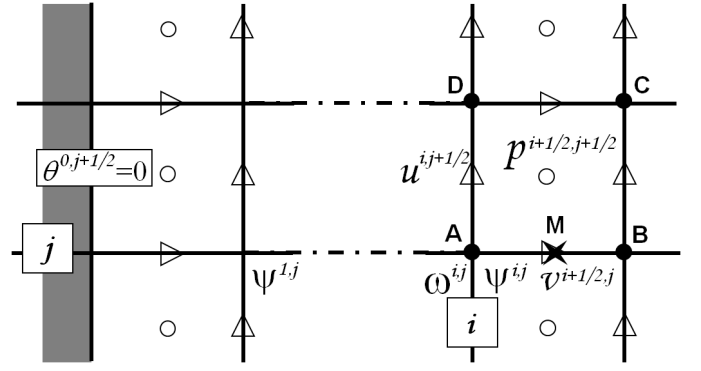}
\caption{MAC grid cells near (left) and away from (right) a boundary} \label{fig:contour}
\end{center}
\end{figure}

Conforming boundaries, which run along the edges of MAC grid cells (see Fig.~\ref{fig:contour}), are treated throughout as a special case of non-conforming ones: the edges which belong to the interface are considered as lying outside the fluid domain $\theta=0$, and the cells they border are, accordingly, deemed irregular.

\subsubsection{Contour integrals around regular cells}
Let us first consider a regular MAC grid cell such as that depicted on Fig.~\ref{fig:contour}-right, and ask how equation~(\ref{integral:p:2d:closed}) for the cell contour $\Gamma=[A,B,C,D,A]$ can be approximated using the discrete values of the fields $p$ and $\mathbf{f}$ at neighboring MAC grid points.

The contribution of e.g. the lower edge $[A,B]$ to $\int_\Gamma\mathbf{n}\cdot\nabla p$ (with $\mathbf{n}$ the outer normal) can be evaluated by Taylor-expanding twice (in the $x$ then $y$ direction) around the midpoint $M$:
\begin{equation}
\label{eqn:integrale:AB}
\begin{split}
-\frac{1}{\Delta x}\int_{A}^{B} p_y(x)\d x &= -\frac{p^{i+\frac{1}{2},j+\frac{1}{2}}-p^{i+\frac{1}{2},j-\frac{1}{2}}}{\Delta y} \\
&+\frac{\Delta y^2}{24}\,p_{yyy}^M - \frac{\Delta x^2}{24}\,p_{yxx}^M\\
&+ O(\Delta x^3,\Delta y^3)
\end{split}
\end{equation}
where e.g. $p_{yxx}^M$ denotes the $yxx$-derivative of the pressure field at $M$.
When this expression is added to its counterpart for the upper edge $[C,D]$, since $-p_{yxx}^M+p_{yxx}^{M'}=O(\Delta x)$, the second order terms partly cancel out and their difference contributes to third order. Adding the contributions from all edges, it finally comes:
\begin{equation}\label{eqn:integrale:ABCDA}
\frac{1}{\Delta x\Delta y}\int_\Gamma \mathbf{n}\cdot\nabla p\,\d s = \Delta^{(d)}p + O(\Delta x^2,\Delta y^2)
\end{equation}

As bulk forces are in principle known analytically, we could compute directly their contribution to the relevant integrals. This however is not too useful in view of the errors introduced by the discretization of the pressure gradients and it is enough to assume, as usual, that the values of the $x$ and $y$ components of bulk forces are discretized on the $u$ and $v$ point (resp.) of the MAC grid. An easy treatment for $\int_\Gamma\mathbf{n}\cdot\mathbf{f}$ then shows:
\begin{equation}\label{eqn:integrale:ABCDA:f}
\frac{1}{\Delta x\Delta y}\int_\Gamma \mathbf{n}\cdot\mathbf{f}\,\d s = \div^{(d)}\mathbf{f} + O(\Delta x^2,\Delta y^2)
\end{equation}

Comparing equations~(\ref{eqn:integrale:ABCDA}) and~(\ref{eqn:integrale:ABCDA:f}) leads to an unsurprising expression:
\begin{equation}
\Delta^{(d)} p=\div^{(d)}\mathbf{f}
\end{equation}
which emphasizes that writing equation~(\ref{integral:p:2d:closed}) around a cell contour is a (convoluted) way to derive a discrete approximation for the Poisson equation.

\subsubsection{Open path integrals around irregular cells}

\begin{figure}[!ht]
\begin{center}
\includegraphics{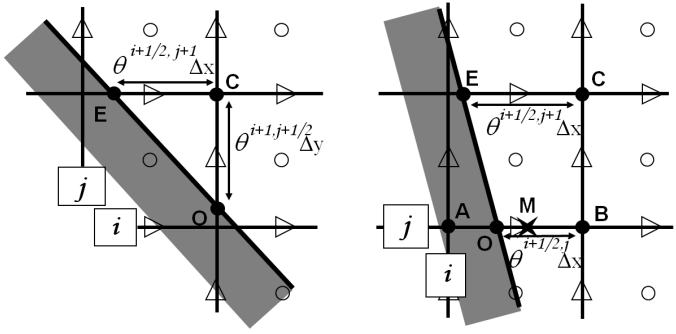}
\caption{Irregular MAC grid cells}
\label{fig:Laplacienpbordirregulier}
\end{center}
\end{figure}

The interest of relying on integral expressions unravels when the Stokes problem is discretized near irregular MAC grid cells, i.e. cells which intersect the boundary, as depicted on Fig.~\ref{fig:Laplacienpbordirregulier}. We then cannot form closed contours lying both on the cell edges and entirely in the fluid domain and must hence introduce open paths, i.e. inject in the problem certain values of $\phi$ at end points $A\ne B$ in the integral expressions of the form~(\ref{integral:p:2d}). The only open paths that can be used must moreover have their end points lying on the interface since this is where the values of $\phi$ are known a priori. We will thus use $\Gamma=[O,B,C,E]$ in the case of Fig.~\ref{fig:Laplacienpbordirregulier}-left and $[O,C,E]$ in that of Fig.~\ref{fig:Laplacienpbordirregulier}-right.

We choose not to introduce extra discretization points, as is done in other implementations~\cite{Russell2003}: the pressure field is only defined at the centers of MAC grid cells. Estimating $\nabla p$ on irregular segments such as $[O,B]$ on Fig.~\ref{fig:Laplacienpbordirregulier}-left or $[O,C]$ on Fig.~\ref{fig:Laplacienpbordirregulier}-right, thus requires using pressure values at the centers of the adjoining cells. Some of these points may lie beyond the interface: we introduce ``ghost'' pressure points.

In the situation illustrated on Fig.~\ref{fig:Laplacienpbordirregulier}-right, two Taylor expansions (in the $x$ and $y$ direction) of $p$ around the midpoint $M$ of edge $[A,B]$ lead to the following estimate:
\begin{equation}\label{eqn:integrale:general}
\begin{split}
\frac{1}{\Delta x}\int_{O}^{B}p_y(x)\d x & = \theta^{i+\frac{1}{2},j}\frac{p^{i+\frac{1}{2},j+\frac{1}{2}}-p^{i+\frac{1}{2},j-\frac{1}{2}}}{\Delta y} \\
& +\theta^{i+\frac{1}{2},j}(1-\theta^{i+\frac{1}{2},j})\frac{\Delta x}{2}\,p^M_{xy}\\
&+ O(\Delta x^2,\Delta y^2)
\end{split}
\end{equation}
which reveals that an error term of order 1 arises as soon as the integration is performed on an irregular ($\theta\ne0,1$) cell edge. The integral $\int\mathbf{n}\cdot\mathbf{f}$ can be analyzed along quite similar lines.

\subsubsection{Discrete error fields}

To lay out the groundwork for our upcoming convergence analysis we define, for each (regular or irregular) edge of the MAC grid, variables $\epsilon_y^{i,j\pm1/2}$ and $\epsilon_x^{i\pm1/2,j}$ that characterize the discretization error. On a horizontal edge $i+\frac{1}{2},j$ (which by definition supports segment $\gamma^{i+\frac{1}{2},j}$):
\begin{equation}\label{eq:error:definition}
\begin{split}
\epsilon_y^{i+\frac{1}{2},j} &=
\frac{1}{|\gamma^{i+\frac{1}{2},j}|}\int_{\gamma^{i+\frac{1}{2},j}}\left(-p_y+f^y\right)\d x\\
&+\frac{p^{i+\frac{1}{2},j+\frac{1}{2}}-p^{i+\frac{1}{2},j-\frac{1}{2}}}{\Delta y}-f^{y\,i+\frac{1}{2},j}
\end{split}
\end{equation}
with the same expression modulo $x,y$-symmetry to define $\epsilon_x^{i,j+\frac{1}{2}}$.
The above Taylor expansions around the segment midpoint showed that:
\begin{equation}\label{eq:epsilonvalue}
\begin{split}
\epsilon_y^{i+\frac{1}{2},j} &= -(1-\theta^{i+\frac{1}{2},j})\frac{\Delta x}{2}\,\left(p^M_{xy}+f^{y,M}_x\right) \\
& + O(\Delta x^2,\Delta y^2)
\end{split}
\end{equation}

For the irregular cell depicted on Fig.~\ref{fig:Laplacienpbordirregulier}-left, using all such expressions to compute the path integral along $\Gamma=[O,B,C,E]$ now leads to:
\begin{equation}
\label{eq:Laplacienpbordirregulier}
\begin{split}
\frac{\phi_{E}-\phi_{O}}{\Delta x\Delta y} &= \theta^{i+\frac{1}{2},j+1}\,\frac{p^{i+\frac{1}{2},j+\frac{3}{2}}-p^{i+\frac{1}{2},j+\frac{1}{2}}}{\Delta y^2}\\
&-\theta^{i+\frac{1}{2},j}\,\frac{p^{i+\frac{1}{2},j+\frac{1}{2}}-p^{i+\frac{1}{2},j-\frac{1}{2}}}{\Delta y^2}\\
&-\frac{p^{i+\frac{1}{2},j+\frac{1}{2}}-p^{i+\frac{3}{2},j+\frac{1}{2}}}{\Delta x^2}-\frac{f^{x,i+1,j+\frac{1}{2}}}{\Delta x}+\frac{\epsilon_x^{i+1,j+\frac{1}{2}}}{\Delta x}\\
&+\frac{\theta^{i+\frac{1}{2},j+1}\,f^{y,i+\frac{1}{2},j+1}-\theta^{i+\frac{1}{2},j}\,f^{y,i+\frac{1}{2},j}}{\Delta y}\\
&-\frac{\theta^{i+\frac{1}{2},j+1}\,\epsilon_y^{i+\frac{1}{2},j+1}-\theta^{i+\frac{1}{2},j}\,\epsilon_y^{i+\frac{1}{2},j}}{\Delta y}
\end{split}
\end{equation}

\subsection{The discrete pressure problem}
\subsubsection{General form and well-posedness}
In our discretization of the pressure problem, the unknowns are the values of $p$ on the set $\Omega_p$ composed of the centers of all regular and irregular cells. An integral equation of the form~(\ref{integral:p:2d}) or~(\ref{integral:p:2d:closed}) is constructed on the contour of each of these cells, thus guaranteeing that the number of equations equals the number of unknowns. Expression~(\ref{eq:Laplacienpbordirregulier}) can be provided a compact form after defining:
\begin{equation}
\left\{
\begin{aligned}
\llbracket \phi \rrbracket &= \frac{\phi_{E}-\phi_{O}}{\Delta x \Delta y} \quad &\text{on irregular cells}\\
\llbracket \phi \rrbracket &= 0 \qquad \qquad &\text{on regular ones}
\end{aligned}
\right.
\end{equation}
where $O$ and $E$ are the origin and end points of the curve $\Gamma$ running counterclockwise between the two intersections of an irregular cell with the interface. With $\theta \nabla^{(d)} p=(\theta \delta_x p, \theta \delta_y p)$, all integral equations constructed using all closed and open paths built using the edges of MAC grid cells can finally be written as:
\begin{equation}\label{operateurPN0}
\llbracket \phi \rrbracket+\div^{(d)}\left(\theta\nabla^{(d)} p\right) = \div^{(d)}\left(\theta f\right)-\div^{(d)}\left(\theta \epsilon\right)
\end{equation}
These equations are so far exact, and dropping the last term leads to the following approximation:
\begin{equation}\label{operateurPN}
\llbracket \phi \rrbracket+\div^{(d)}\left(\theta\nabla^{(d)} p\right) = \div^{(d)}\left(\theta f\right)
\end{equation}
which is our discretized pressure problem: the only unknowns left are now just the values of $p$ on $\Omega_p$. This numerical scheme, solves~(\ref{operateurPN0}) with additional sources $=\div^{(d)}\left(\theta \epsilon\right)$, which hence accounts for all the discretization errors.

The discrete form~(\ref{operateurPN}) of the pressure problem presents several niceties. First, it is quite simple to implement, as it involves nearly usual discretized forms of the divergence operator. Second, it guarantees that the discrete pressure problem is well-posed: indeed, the underlying consistency condition between sources and boundary conditions (the discrete form of $\int_{\partial\Omega}\mathbf{n}\cdot\mathbf{f}=\int_\Omega\div\mathbf{f}$), which is required to ensure that the discrete problem has a solution, is by construction verified on each cell, hence on the full problem. Third, its matrix representation is symmetric, thus allowing the use of fast algorithms.

The $O(1)$ errors present in $\div^{(d)}\left(\theta \epsilon\right)$ near irregular cells, however, seem foreboding and the outstanding issue is now to analyze their consequences for the pressure solution.

\subsubsection{Convergence analysis}
\label{sec:error:p}
Our examination of the numerical errors in the resulting pressure approximation borrows from the argument developed by Jomaa and Macaskill~\cite{JomaaMacaskill2005} and Gibou and Fedkiw~\cite{GibouFedkiwCheng2002} to analyze the discretization of the Poisson-Dirichlet problem at irregular cells in similar finite-difference schemes. They have observed that despite the introduction of an error of order 0 in the discretization of the Laplacian, the solution for the Dirichlet-Poisson problem could still be obtained with second order accuracy. The reason is that certain discretization errors can be mapped into higher order errors in the boundary condition.

To disentangle, from the discrete field $\mathbf{\epsilon}=(\epsilon_x,\epsilon_y)$, an effective error on the boundary conditions of the pressure problem, we now seek to write:
\begin{equation}\label{eq:helm:epsilon}
\llbracket \chi \rrbracket+\div^{(d)}\left(\theta\nabla^{(d)} q\right) = \div^{(d)}\left(\theta \epsilon\right)
\end{equation}
where $q$ is a discrete field with values on $\Omega_p$ (the same points as $p$) and $\chi$ is defined on $\Omega'_\phi$, the set of all irregular end points of irregular segments (these points, of course, lie on the boundary).

Let $\widetilde\Omega_p\subset\Omega_p$ denote the set of all the centers of regular cells. The discrete Poisson problem:
\begin{subequations}
\begin{align}\label{eq:poisson:q}
\Delta^{(d)}q &= \nabla\cdot^{(d)}(\theta\epsilon)&\text{on}\quad\widetilde\Omega_p\\
q&=0 &\text{on}\quad\Omega_p\setminus\widetilde\Omega_p\label{eq:discrete:poisson:dirichlet}
\end{align}
\end{subequations}
defines a unique discrete field $q$ with values on $\Omega_p$. As~(\ref{eq:poisson:q}) is written on regular cells only, all $\theta$ values appearing under the divergence are actually equal to 1, and $\Delta^{(d)}q=\div^{(d)}\left(\theta\nabla^{(d)} q\right)$ at these points. Furthermore (see the discussion preceding equation~(\ref{eqn:integrale:ABCDA})), the source term in the above Poisson problem, $\nabla\cdot^{(d)}(\theta\epsilon)=\nabla\cdot^{(d)}(\epsilon)=O(\Delta x^2)$: the solution $q$ is therefore also of order $O(\Delta x^2)$.

On regular cells, the field $\mathbf{\epsilon}-\nabla^{(d)} q$ is, by construction, divergence-free, hence can be written as a discrete curl. In other words, there is a scalar field $\chi$ defined on the set $\widetilde\Omega_\phi$ of all corners of regular cells, such that for each regular segment $\gamma$:
\begin{equation}\label{eq:chi}
\epsilon^\gamma=\left[\nabla^{(d)} q\right]^\gamma+\frac{\chi_B-\chi_A}{|\gamma|}
\end{equation}
with $|\gamma|$ the segment length, $A$ and $B$ (with $B>A$ using the ordering naturally inherited from the $x$ and $y$ coordinates) its two end points. Indeed, in perfect analogy with the continuum construction of a potential field, $\chi$ can be defined by considering paths $\Gamma$ made of regular segments; for $\Gamma$ with end points $A$ and $B$:
\begin{equation}
\chi_B-\chi_A=\sum_{\gamma\subset\Gamma}\overline\gamma\,\left(\epsilon^\gamma-\left[\nabla^{(d)} q\right]^\gamma\right)
\end{equation}
where $\overline\gamma=\pm|\gamma|$ is a signed scalar accounting for the direction in which $\Gamma$ runs along segment $\gamma$. $\chi$ is well-defined up to a constant because equation~(\ref{eq:poisson:q}) guarantees that for any closed loop, the rhs vanishes in the above equation. If we now consider two points $C$ and $D$ at a fixed physical distance, the difference $\chi_C-\chi_D$ can be constructed using paths containing $O(1/\Delta x)$ segments; each term in the sum in of order $O(|\gamma|) \times O(\epsilon) = O(\Delta x^3)$, whence $\chi=O(\Delta x^2)$ (it is so far defined on the corners of regular cells).

Requiring additionally that Eq.~(\ref{eq:chi}) holds on all irregular segments now defines $\chi$ on $\Omega'_\phi$. On each irregular segment $\gamma'$, the Dirichlet condition~(\ref{eq:discrete:poisson:dirichlet}) ensures that $\left[\nabla^{(d)} q\right]^{\gamma'}=0$ while $\epsilon^{\gamma'}=O(\Delta x)$ (this was the most problematic error): the constructed values of $\chi$ on $\Omega'_\phi$ are hence again $O(\Delta x^2)$.

With these definitions, it is easy to check that Eq.~(\ref{eq:helm:epsilon}) holds. Plugging it into equation~(\ref{operateurPN0}), now yields an equivalent set of exact equations:
\begin{equation}\label{operateurPN2}
\llbracket \phi-\chi \rrbracket+\div^{(d)}\left(\theta\nabla^{(d)} (p-q)\right) = \div^{(d)}\left(\theta\mathbf{f}\right)
\end{equation}
which shows that the discretization errors introduced when solving~(\ref{operateurPN}) can be separated into two contributions: (i) the usual bulk error $q$ due to the discretization of the Laplacian on regular cells; (ii) a boundary error which perturbs $\phi$ at second order. This latter error contributes at most second order errors in the underlying Poisson-Dirichlet problem defining $\phi$, hence in the solution of the pressure problem. All the discretization errors affecting the resulting pressure approximation hence arise at second order.

\subsection{Numerical validation}\label{sec:test:pressure}

\begin{figure}[!ht]
\begin{center}
\includegraphics{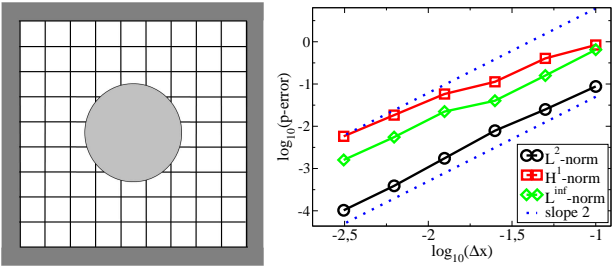}
\includegraphics{fig1.eps}
\caption{Left: The domain used for our numerical test: a square domain $\Omega=[0.4,2.4]^2$ with a cylinder of radius $1/6$. Right: the results of our convergence analysis.}\label{fig:castestshyb}
\end{center}
\end{figure}

To validate our implementation, we will use throughout this paper a test Stokes problem introduced in~\cite{GuermondMinevShen2006}. The time-varying velocity and pressure fields are chosen to be:
\begin{subequations}
\label{eq:testcase}
\begin{align}
u(x,y) &= \pi\,\sin(t)\,\sin^2(\pi x)\,\sin(2\pi y)\\
v(x,y) &= \pi\,\sin(t)\,\sin(2\pi x)\,\sin^2(\pi y)\\
p(x,y) &= 4\pi^2\,\sin(t)\,\cos(\pi x)\,\sin(\pi y)
\end{align}
\end{subequations}
and the body forces are computed to match Eq.~(\ref{eqn:Stokes}).
We use the integration domain described in Fig.\ref{fig:castestshyb}-left: $\Omega=[0.4,2.4]^2\setminus I$ where $I$ is an inclusion of radius $r=1/6$ located near the domain center. Using $[0.4,2.4]^2$ as basis of the integration domain permits avoiding trivial error cancellations due to the symmetry of the solution. The inclusion $I$ is also slightly off-center with respect to $[0.4,2.4]^2$ so as to avoid other possible error cancellations due to symmetries with respect to the grid.

The convergence of our solver for the pressure problem is tested using~(\ref{eq:testcase}) at time $t=1$. The analytical boundary values are then:
\begin{equation}
\begin{split}
&\phi(x,y) =\\
&2 \pi^2 \text{sin}(1) \left(\text{cos}(2 \pi x) \text{sin}^{2}(\pi y)
+ \text{cos}(2 \pi y) \text{sin}^{2}(\pi x)\right) \\
& - \text{cos}(1)\text{sin}(2 \pi y)\text{sin}^{2}(\pi x)
\quad.
\end{split}
\end{equation}
Several resolutions $N_x\times N_y$ are used, with $N_x=N_y=20, 40, 80, 160, 320, 640$, corresponding to discretization steps $\Delta x = L_x/N_x$ and $\Delta y = L_y/N_y$, with cell dimensions $L_x=L_y=2$.
The matrix representing the discrete problem is sparse, allowing us to use band storage; it is also symmetrical, but we have nevertheless used a bi-conjugate gradient stabilized algorithm (BICGSTAB) preconditioned by ILU factorization which we had been using throughout this study for its robustness.

The test results presented in figure~\ref{fig:castestshyb}-right clearly show second order for ${L^2}$, ${L^\infty}$, and ${H^1}$ norms, thus validating the convergence analysis performed in the previous section.

\section{Implementation of the Stokes problem}
\label{sec:implementationStokesPb}

We now turn to the actual implementation of the unsteady Stokes problem~(\ref{eqn:Stokes}): as our focus remains on the details of the spatial discretization, we prefer to keep the time-integration as simple as possible and will consider only explicit schemes. Higher-order-in-time and implicit schemes based on the ideas developed in the present paper are left for future works.

With $\Delta t$ the time-step, and denoting $u^n$, $v^n$, etc. the values of hydrodynamic fields at $t^n=n\,\Delta t$, the semi-discrete Euler scheme for the Stokes problem simply reads:
\begin{subequations}
\label{eq:Stokes:n}
\begin{align}
\label{eq:Stokes:n0}
\rho\left(\frac{\mathbf{u}^{n+1}-\mathbf{u}^n}{\Delta t}\right)  & =  -\nabla p - \mu\nabla^\perp{\omega^n} + \mathbf{f}(t^n)\\
\div\mathbf{u}^{n+1} &= 0\\
\mathbf{u}^{n+1}&=\mathbf{u}(t^{n+1}) \quad\text{on}\quad\partial\Omega
\end{align}
\end{subequations}
where we note $\mathbf{u}(t^{n+1})$ (resp. $\mathbf{f}(t^n)$) the (exact) boundary values of the velocity (resp. bulk forces) to be distinguished from the approximated solution $\mathbf{u}^{n+1}$.

To fully exploit our above analysis, we write the corresponding pressure problem as:
\begin{equation}\label{eq:pressure:continuous}
\begin{split}
\nabla^\perp\left(\rho\left(\frac{\psi(t^{n+1})-\psi^{n}}{\Delta t}\right) + \mu\omega^n \right) +\nabla p =  \mathbf{f}(t^n)
\end{split}
\end{equation}
The notation $\psi(t^{n+1})$ emphasizes that only exactly known boundary values are introduced via this term: indeed, from the above analysis (see the structure of Eq.~(\ref{operateurPN0}) and~(\ref{operateurPN}), or the integral form~(\ref{integral:p:2d}) they derive from) we can anticipate that all terms under the $\nabla^\perp$ operator enter the pressure problem only via their boundary values. We note that accordingly, boundary values of both $\psi^{n}$ and $\omega^n$ will have to be estimated from the discrete velocity field $\mathbf{u}^n$.

\subsection{Spatial discretization}
\label{sec:discretization}

Following Section~\ref{sec:pressure} the pressure problem is discretized as:
\begin{equation}\label{eq:pressure:discrete0}
\left\llbracket \rho\left(\frac{\psi(t^{n+1})-\psi^{n}}{\Delta t}\right) + \mu\omega^n \right\rrbracket+\nabla\left(\theta\nabla^{(d)} p\right) =
\div^{(d)}\left(\theta\mathbf{f}\right)
\end{equation}
while introducing only second order errors. The term $\llbracket\psi^{n}\rrbracket/\Delta t$ and its relationship with the velocity field can immediately be clarified by introducing the discrete currents $\mathbf{\bar u}^n$, localized at the same MAC grid points as the corresponding velocities:
\begin{align*}\label{eq:current:definition}
\left({\bar u}^{n}\right)^{i,j+\frac{1}{2}} &= \frac{1}{|\gamma^{i,j+\frac{1}{2}}|}\int_{\gamma^{i,j+\frac{1}{2}}} u^n(x,y)\,\d x\\
\left({\bar v}^{n}\right)^{i+\frac{1}{2},j} &= \frac{1}{|\gamma^{i+\frac{1}{2},j}|}\int_{\gamma^{i+\frac{1}{2},j}} v^n(x,y)\,\d y
\end{align*}
This definition justifies the introduction of $\mathbf{\bar u}^n$ values at ``ghost'' grid points, beyond the boundary, where needed to account for the mass current passing through irregular segments.

With these definitions, the following relation:
\begin{equation}\label{eq:curl:exact}
\llbracket\psi^n\rrbracket = \div^{(d)}\left(\theta\mathbf{\bar u}^n\right)
\end{equation}
is exact, and equation~(\ref{eq:pressure:discrete0}) is immediately recast as:
\begin{equation}\label{eq:pressure:discrete}
\left\llbracket \frac{\rho\psi(t^{n+1})}{\Delta t} + \mu\omega^n \right\rrbracket+\nabla\left(\theta\nabla^{(d)} p\right) =
\div^{(d)}\left(\theta\left(\mathbf{f}+\frac{\rho\mathbf{\bar u}^n}{\Delta t}\right)\right)
\end{equation}
This discrete problem approximates the set of all integral equations~(\ref{integral:p:2d}) around the cells' contours, and requires the boundary values of the fields under $\llbracket.\rrbracket$ to be estimated at second order on all the intersections of the MAC grid with the interface. It provides the values of $p$ at the centers of all regular and irregular cells, which may include a few ghost points in the latter case. Accordingly, the gradient of pressure as appearing on the rhs of (\ref{eq:Stokes:n0}), can be estimated from the p values, using only centered differences, at all the velocity points that lie on either regular or irregular edges (these again, may include a few ghost velocity points).

Our spatial discretization of the Stokes problem~(\ref{eq:Stokes:n}) relies on the following principles:
\begin{itemize}
\item We only use values of the fields $u$, $v$, and $p$ at their usual grid points, i.e. at the centers of resp. edges and cells.
\item We only use centered differences to estimate the rhs of Eq.~(\ref{eq:Stokes:n}) so as to introduce only second order errors in the evaluation of these terms.
\end{itemize}
We recognize that introducing extra discretization points or more extended stencils in the estimation of these terms might be interesting strategies, but it is quite different in spirit from what we are developing here, and we therefore restrict to the above rules. The second requirement implies that to estimate the term $\nabla^\perp\omega$, we need to define the vorticity field on all regular end points of all edges on which the velocity field is defined. These again may involve a few ghost values.

Assuming that the values of $\omega$ and $p$ are defined on the specified points, it is clear that Eq.~(\ref{eq:Stokes:n0}) can be iterated once. With this convention for the discrete points where $\omega^n$ is defined, we can further rewrite equation~(\ref{eq:pressure:discrete}) as:
\begin{equation}\label{eq:pressure:discrete:final}
\begin{split}
\left\llbracket \frac{\rho\psi(t^{n+1})}{\Delta t}\right\rrbracket+\nabla\left(\theta\nabla^{(d)} p\right) =&\\
\div^{(d)}&\left(\theta\left(\mathbf{f}+\frac{\rho\mathbf{\bar u}^n}{\Delta t}+\mu\nabla^\perp\omega^n \right)\right)
\end{split}
\end{equation}
indeed, it is easy to check that evaluating the boundary values of $\omega^n$ in~(\ref{eq:pressure:discrete}) by standard (second order) linear interpolation on irregular segments, is strictly equivalent (numerically) to introducing the term $\div^{(d)}\left(\theta\nabla^\perp\omega^n \right)$.

Our discretized numerical problem is finally subsumed as Eq.~(\ref{eq:Stokes:n0}) and~(\ref{eq:pressure:discrete:final}). The remaining difficult questions are: given $u$ and $v$ on regular and irregular cell edges at time $t^n$, how can the corresponding pressure problem be defined? how can the value of $\omega$ be determined at all appropriate points? how can the discrete fluxes $\mathbf{\bar u}^n$ be computed at the required order of approximation (and what is the required order of approximation)?

\subsection{Vorticity estimates}

\begin{figure}[!ht]
\begin{center}
\includegraphics{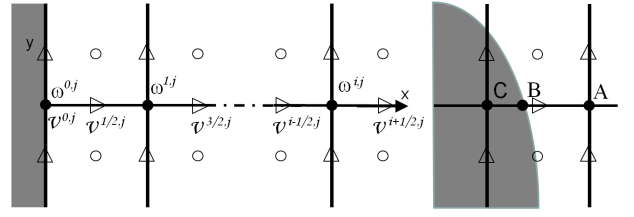}
\caption{Localization of the velocity and vorticity fields: left, near a regular interface; center, at a regular grid point; right, near a irregular interface.}\label{fig:extrapolationvwmur}
\end{center}
\end{figure}

As usual, the vorticity $\omega=u_y-v_x$ at a regular point can be estimated using centered differences (see illustration on Fig.~\ref{eq:pressure:discrete:final}-center):
\begin{equation}\label{eq:discrete:rot}
\omega=\rot^{(d)}\mathbf{u}+O(\Delta x^2)=\delta^+_x v-\delta_y^+ u+O(\Delta x^2)
\end{equation}

Let us now consider the case of a regular interface, as depicted on Fig.~\ref{fig:extrapolationvwmur}-left.
In this example of a vertical wall, $u$ and all its $y$-derivatives, especially $u_y$, are known along the boundary: estimating $\omega$ at these points only requires to introduce an approximation for $v_x$. This is performed by Taylor expanding $v$ in the $x$ direction: taking the abscissa of the vertical wall as origin of $x$ coordinates and $i$ indices, it comes:
\begin{equation}\label{dl2}
v_x^{0,j}
 = \frac{1}{3\Delta x}\left(-8\,v^{0,j}+9\,v^{\frac{1}{2},j} - v^{\frac{3}{2},j}\right)
+ O(\Delta x^2)
\end{equation}
This expression involves only discrete values of $v$ plus its boundary value $v^{0,j}$.

The resulting second order approximations\footnote{An alternative way to obtain a second order approximation for $\omega$ at this regular boundary is to use a linear extrapolation from the values of $\omega$: $\omega^0=2\omega(\Delta x)-\omega(2\,\Delta x)+O(\Delta x^2,\Delta y^2)$.
The sought second order accuracy is obtained provided the bulk value of $\omega$ are known at this order, which can easily be done at regular grid points.}
for $\omega = u_y-v_x$ play in our implementation a role analogous to Thom's formula and its many variants~\cite{ELiu1995,Thom1933}. However, as we do not work in a vorticity-potential formulation, it is the values of the velocity field which appear in the rhs.

We illustrate on Fig.~\ref{fig:extrapolationvwmur}-right several situations which are encountered when evaluating $\omega$ near an irregular boundary:
\begin{itemize}
\item $A$ lies within $\Omega$;
\item $B$ lies on the interface;
\item $C$ is a ghost point.
\end{itemize}
At all these points, $\omega$ must be estimated at second order. The velocity fields $u$ and $v$ are defined at the centers of both regular and irregular segments emanating from $A$, hence $\omega^A$ can be simply computed using the regular expression~(\ref{eq:discrete:rot}), which is second order accurate. At point $B$, we can obtain second order accuracy using a classical linear interpolation expression involving $\omega^C$ and $\omega^A$, assuming that $\omega^C$ is second order accurate. The outstanding question is how to evaluate $\omega$ with second order accuracy at all ghost points?

\begin{figure}[!ht]
\begin{center}
\includegraphics{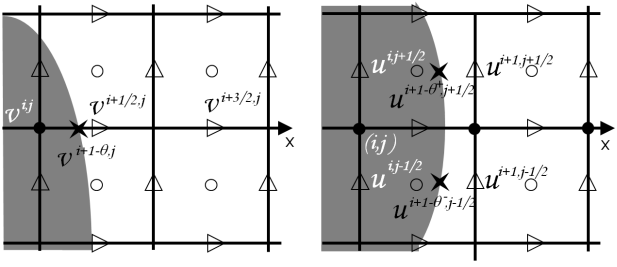}
\caption{Localization of velocity field $v$ and vorticity field $\omega$ near an irregular interface on the Mac grid} \label{fig:interface}
\end{center}
\end{figure}

Let us first consider the case depicted on Fig.~\ref{fig:interface}-left. To avoid singularities due to the possible mismatch between the discrete value $v^{i+\frac{1}{2},j}$ and the analytical boundary value $v^{i+1-\theta^{i+\frac{1}{2},j},j}$, we need to construct approximations which involve the velocity at points that are away from the interface. It is convenient to write a general expression based on points $({i+n+\frac{1}{2},j})$, and $({i+n+\frac{3}{2},j})$. Using a Taylor expansion of $v$ along the direction $x$ with origin at the ghost point, we find:
\begin{equation}\label{eq1:LN2grain}
\begin{split}
v_x^{i,j} &=\frac{1}{\eta\Delta x}\left(\alpha\,v^{i+1-\theta^{i+\frac{1}{2},j},j}-\beta\,v^{i+n+\frac{1}{2},j} + \gamma\,v^{i+n+\frac{3}{2},j}\right)\\
&+ O(\Delta x^2)
\end{split}
\end{equation}
with:
\begin{equation}\label{eq1:params}
\begin{split}
\alpha&=8(n+1)\\
\beta &=4\left((1-\theta^{i+\frac{1}{2},j})^2-\left(n+\frac{3}{2}\right)^2\right)\\
\gamma &=-4\left((1-\theta^{i+\frac{1}{2},j})^2-\left(n+\frac{1}{2}\right)^2\right)\\
\eta   &=\alpha\,(1-\theta^{i+\frac{1}{2},j})+\beta\,\left(n+\frac{1}{2}\right)+\gamma\,\left(n+\frac{3}{2}\right)
\end{split}
\end{equation}
We note that in the case $n=0$, this expression is identical to Eq.(\ref{dl2}). Near an irregular interface, it will be used with $n=1$.

This approximation for $v_x$ and its counterpart for $u_y$ suffice to define $\omega$ at the ghost point depicted on Fig.~\ref{fig:interface}-left. However, few cases also exist when a ghost point lies at the end of a single irregular segment, as illustrated on Fig.~\ref{fig:interface}-right. It this latter situation, it is not possible to use~(\ref{eq1:LN2grain}) to estimate $u_y$. This is performed by first computing ghost values of $u$ at the points $i,j\pm\frac{1}{2}$ via Taylor expansion. There again, the mismatch between discrete values $u^{i+1,j\pm\frac{1}{2}}$ and nearby boundary values $u^{i+1-\theta^\pm,j\pm\frac{1}{2}}$ may bring in singularities which we avoid by using discrete $u$ values which are away from the interface. We thus use:
\begin{equation}\label{eq2:LN2grain}
\begin{split}
u^{i,j\pm\frac{1}{2}} &=\frac{1}{\eta\Delta x}\left(\alpha\,u^{i+1-\theta^\pm,j\pm\frac{1}{2}}-\beta\,u^{i+2,j\pm\frac{1}{2}}+ \gamma\,u^{i+3,j\pm\frac{1}{2}}\right)\\
&+ O(\Delta x^2)
\end{split}
\end{equation}
with $\theta^\pm = \theta^{i,j\pm\frac{1}{2}}$ and:
\begin{equation}\label{eq2:params}
\begin{split}
\alpha&=4\\
\beta &=2\theta^{i,j\pm\frac{1}{2}}\left(\theta^{i,j\pm\frac{1}{2}}-3\right)\\
\gamma &=\frac{4}{3}\theta^{i,j\pm\frac{1}{2}}\left(2-\theta^{i,j\pm\frac{1}{2}}\right)\\
\eta   &=\alpha+\beta+\gamma
\end{split}
\end{equation}
Using $u^{i,j\pm\frac{1}{2}}$, then $u_y^{i,j}$ is immediately obtained by centered differences, which finishes to define $\omega$ on ghost points such as that illustrated on Fig.~\ref{fig:interface}-right.

\subsection{Discrete fluxes}

The discrete fluxes $\mathbf{\bar u}^n$ appear in the pressure problem via the term $\div^{(d)}\left(\theta\frac{\mathbf{\bar u}^n}{\Delta t}\right)$. To understand what order of approximation is required in the evaluation of these fluxes, let us recall that the discretization errors of the pressure problem are entirely contained in a term of the form $\div^{(d)}\left(\theta\epsilon\right)$: we found in Section~\ref{sec:pressure} that second order accuracy on pressure is achieved even though $\epsilon$ may present first order errors on irregular segments. Consequently, $\frac{\mathbf{\bar u}^n}{\Delta t}$ may present first order errors on irregular segments (it must of course be second order accurate on regular ones). Within our time-explicit scheme, as the CFL condition entails that $\Delta t\sim\Delta x^2$, $\mathbf{\bar u}^n$ must be computed with fourth and third order accuracy on resp. regular and irregular segments.

The previous discussion concerning the extrapolation of $\omega^n$ to boundary and ghost points relies on the evaluation of both $u_y$ and $v_x$ at the required points. It turns out to be convenient to use these intermediary fields in the estimation of the discrete fluxes at the appropriate order, via expression of the form:
\begin{equation}\label{eq:expansion:u}
\begin{split}
\bar v^{i+\frac{1}{2},j} &= v^{i+\frac{1}{2},j}+\frac{\alpha\Delta x}{2}\,\left(v_x^{i,j}+v_x^{i+1,j}\right)\\
&+{\beta\Delta x^2}\,\left(v_x^{i+1,j}-v_x^{i,j}\right) +O(\Delta x^3)
\end{split}
\end{equation}
where:
\begin{equation}\label{eq:params}
\begin{split}
\alpha&=\frac{1-\theta^{i+\frac{1}{2},j}}{2}\\
\beta &=\frac{1}{6}\left(\theta^{i+\frac{1}{2},j}\right)^2-\frac{1}{4}\theta^{i+\frac{1}{2},j}+\frac{1}{8}
\end{split}
\end{equation}

This finishes to define our discrete problem.

\subsection{Numerical validation}

\begin{figure}[!ht]
\begin{center}
\subfigure[Error on $\omega$ field]{\epsfig{figure=fig2.eps}}
\subfigure[Error map on $\omega$]{\epsfig{figure=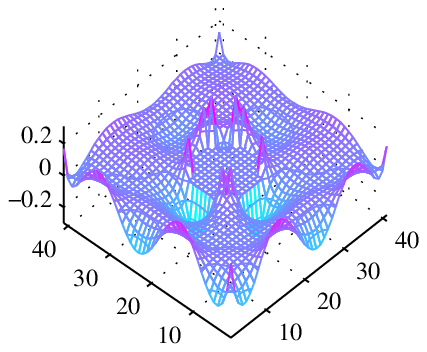}}
\subfigure[Error on $u$]{\epsfig{figure=fig4.eps}}
\subfigure[Error map on $u$]{\epsfig{figure=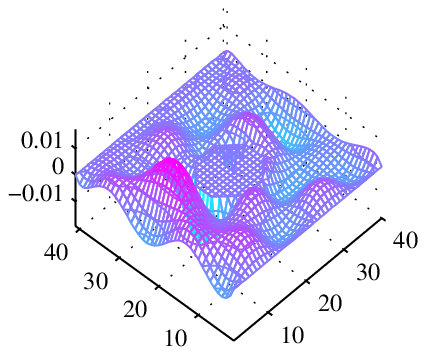}}
\subfigure[Error on $p$]{\epsfig{figure=fig6.eps}}
\subfigure[Error map on $p$]{\epsfig{figure=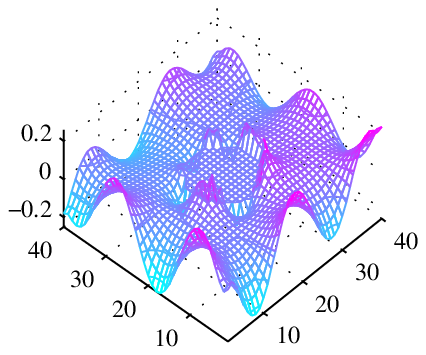}}
\subfigure[Error on divergence]{\epsfig{figure=fig8.eps}}
\subfigure[Error map on divergence]{\epsfig{figure=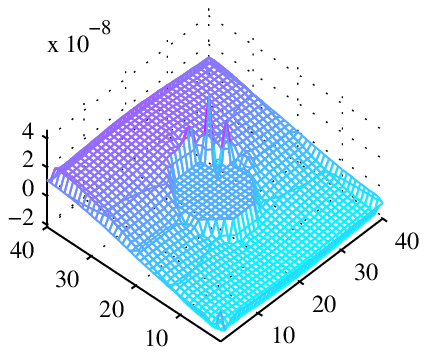}}
\end{center}
\caption{From top to bottom: errors on $\omega$, $u$, $p$, and $\div\mathbf{u}$. Left: convergence analysis. Right: error maps on a 40$\times$40 system.}
\label{fig:validation}
\end{figure}

We are now in position to implement the numerical scheme defined by equations~(\ref{eq:Stokes:n}) and~(\ref{eq:pressure:discrete:final}) modulo the above technicalities in the computation of $\omega^n$ and $\mathbf{\bar{u}}^n$ near and at boundaries. To do this, we implement the time-dependent test case described in Section~\ref{sec:test:pressure}. Panels (a,c,e) on Fig.~\ref{fig:validation}-left present the convergence analysis in norms $L^2$, $H^1$ and $L^\infty$, when averaged over time during numerical integration up to time $t=1$. These norms are computed using either the set of all regular grid cells (dashed lines) or using both regular and irregular grid cells (solids lines), which may include a few ghost points. A clear second order convergence is thus found for all the norms and fields considered. A few maps of the error field are also provided on the right of Fig.~\ref{fig:validation}, panels (b,d,f).

To validate our algorithm, we also need to check that mass conservation
$$\left\llbracket \psi(t^n)\right\rrbracket=\div^{(d)}(\theta\mathbf{\bar u}^{n})$$
is properly upheld. This is done by computing the values $\div^{(d)}(\theta\mathbf{\bar u}^{n})-\left\llbracket \psi(t^n)\right\rrbracket$ or regular and irregular cell (see Fig.~\ref{fig:validation}-g), which shows faster than second order convergence. We also show on the same panel that $\div^{(d)}(\theta\mathbf{u}^{n})-\left\llbracket \psi(t^n)\right\rrbracket$ on regular cells presents only round-off errors, in the range of $10^{-8}$.

\begin{figure}[!ht]
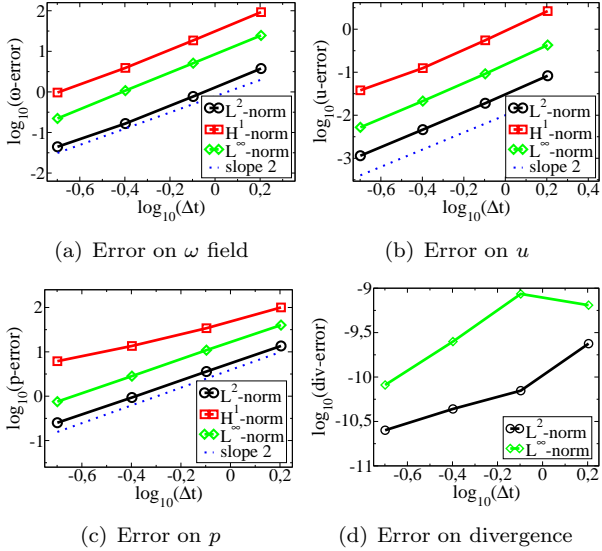

\begin{center}
\subfigure[Error on $\omega$ field]{\epsfig{figure=fig10.eps}}
\subfigure[Error on $u$]{\epsfig{figure=fig11.eps}}
\subfigure[Error on $p$]{\epsfig{figure=fig12.eps}}
\subfigure[Error on divergence]{\epsfig{figure=fig13.eps}}
\end{center}
\caption{Convergence analysis on $\omega$, $u$, $p$, and $\div\mathbf{u}$ on a 40$\times$40 system.}
\label{fig:convergence:time}
\end{figure}

The above analysis was based on the Euler scheme~(\ref{eq:Stokes:n}) and~(\ref{eq:pressure:discrete:final}), which can at most provide first order convergence in time. To complete our analysis, we thus also consider a standard mid-point scheme based on this Euler step. Convergence in time is then tested using $\mu=10^{-1}$, $\rho=1000$, with a fixed space discretization $N_x=N_y=40$, i.e. $\Delta x=\Delta y=5\times10^{-2}$ and by varying $\Delta t=0.2,0.4,0.8,1.6$. The results presented in Fig.~\ref{fig:convergence:time} show again a clear second order convergence.

\section{Test cases}
\label{sec:testcases}
\subsection{The Fax\'en problem}

\begin{figure}[!ht]
\begin{center}
\includegraphics{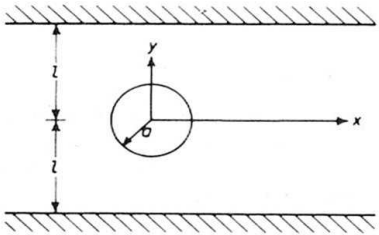}
\caption{A cylinder of radius $a$ moving at a constant speed $\mathbf{U}=U\,\mathbf{e}_x$ along the median axis of an infinite horizontal channel of width $2\ell$} \label{fig:faxencylindre}
\end{center}
\end{figure}

To further test our algorithm, we now consider the case depicted on Fig.~\ref{fig:faxencylindre} of a cylinder of radius $a$ moving at a constant speed $\mathbf{U}=U\,\mathbf{e}_x$ along the median axis of an infinite horizontal channel of width $2\ell$. We focus on steady state, when several approximations exists for the drag coefficient $C$:
Fax\'en~\cite{Faxen1946} provides an asymptotic expression for $C$ in the limit $k=a/\ell\ll1$:
\begin{equation}\label{Cfaxenana}
\begin{split}
C(k) &= \frac{F_x(k)}{\mu U} \\
&= \frac{4\pi}{A_0-\ln(k)+ A_{2}k^2+ A_{4}k^4+ A_{6}k^6+ A_{8}k^8}\\
\end{split}
\end{equation}
\begin{equation}
\text{with }\left\{
\begin{array}{lll}
A_0 &=& -0.9156892732 \\
A_2&=&1.7243844 \\
A_4&=&-1.730194\\
A_6&=&2.405644\\
A_8&=&-4.59131
\end{array}
\right.
\end{equation}
The reverse limit, $a/\ell\to1$ has been studied by Bungay and Brenner~\cite{BungayBrenner1973}: using $\epsilon =(1-k)/k$, they find:
\begin{equation}\label{lubritot}
\begin{split}
C(\varepsilon)&= 9\pi \sqrt{2} \varepsilon^{-5/2} + 24 B \varepsilon^{-2} + 6\pi \sqrt{2} \varepsilon^{-3/2}\\
&+ (24C + 12D)\varepsilon^{-1}+ 2\pi \sqrt{2} \varepsilon^{-1/2} + \ldots \\
\end{split}
\end{equation}
where $B$, $C$, and $D$ are integration constants.

We here use these asymptotic expansions as benchmarks tests for our implementation. To avoid difficulties due to the progression of the cylinder with respect to the gridding, we assume the grid in fixed in the frame of the cylinder: the problem is identically mapped onto the case of a fixed cylinder, in a channel with moving walls, where $\mathbf{u}=-\mathbf{U}$. The pressure gradient along the channel -- implemented via homogeneous bulk forces along $x$ -- is computed at each timestep to enforce the condition that the average fluid velocity through is also $=-\mathbf{U}$.

The drag force is in principle given by a path integral involving the contour encircling the cylinder (see Fig.~\ref{fig:faxencylindre} for details):
\begin{equation}
F=\oint_\Gamma\,\Pi\mathbf{n}\d s
\end{equation}
with the hydrodynamic stress tensor:
\begin{equation}\label{Pi}
\Pi=\begin{pmatrix}
-p+2\mu\frac{\partial u}{\partial x}& \displaystyle{\mu\left(\frac{\partial u}{\partial y}+\frac{\partial v}{\partial x}\right)}\\
\displaystyle{\mu\left(\frac{\partial u}{\partial y}+\frac{\partial v}{\partial x}\right)}&-p+2\mu\frac{\partial v}{\partial y}
\end{pmatrix}
\end{equation}
As we focus on steady state -- which we reach after a reasonably short transient -- momentum conservation $\div\Pi=0$ guarantees that $F$ can be identically computed on any contour $\Gamma$ encircling the cylinder: it enables us to use contours based on the regular segments of the grid so as to avoid introducing extraneous errors due to the estimation of $\Pi$ along the boundary and only catch errors coming from the simulation proper. We have checked that using different contours provides identical results modulo round-off errors.

\begin{figure}[!ht]
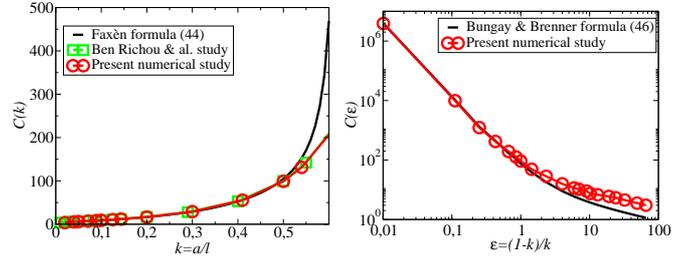

\begin{center}
\includegraphics{fig14.eps}
\includegraphics{fig15.eps}
\caption{Comparison between numerical and asymptotic values of the drag coefficient $C$ for a cylinder driven in an infinite channel. Left: the Fax\'en case $k=a/\ell\ll1$. Right: the lubrification regime $(k\rightarrow 1)$} \label{fig:FaxenResults}.
\end{center}
\end{figure}

The simulation is implemented, following Ben Richou~{\it et al\/}~\cite{Benrichou2005}, using a constant radius $a=0.8$mm, and a fixed spatial step $\Delta x=\Delta y=2.0\times10^{-4}$, while varying the channel width and accordingly the number of grid points $N_x\times N_y$. The CFL condition requires the integration time-step to be of order $5.10^{-3}$. Our measurements of the drag coefficient, presented on Fig.~\ref{fig:FaxenResults}, do converge towards the relevant asymptotic approximation in both limits.

\subsection{Flow through a porous medium}

\begin{figure}[!ht]
\label{fig:porous}
\begin{center}
\includegraphics{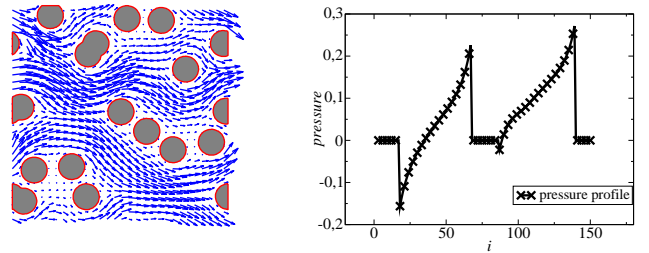}
\includegraphics{fig17.eps}
\caption{Flow through a porous media. Left: arrow map of the velocity field. Right: pressure profile along a horizontal cut through the middle of the simulation cell.} \label{fig:Porousmedia}
\end{center}
\end{figure}

To further illustrate the capabilities of our algorithm, we have implemented a Stokes flow through an array of randomly placed cylinder, as illustrated on Fig.~\ref{fig:porous}, which is a simplified 2D model of a porous medium. Cuts of the pressure field along the $x$ direction show clear singularities near the obstacles, showing they are very well captured.

\section{Extension to 3D}
\label{sec:conclusion:3D}

Let us now come back to the case of a 3D flow, as discussed at the beginning of Section~\ref{sec:helmholtz3d}: writing the velocity field as $\mathbf{u}=\rot\mathbf{A}$, with $\mathbf{A}$ a potential vector, the Stokes equation reads:
\begin{equation}
\rho\,\rot\mathbf{A}_t+\mu\,\rot\mathbf{\omega} + \nabla p = \mathbf{f}
\end{equation}
which can be viewed as the following Helmholtz decomposition:
\begin{equation}
\label{eq:h}
\rot\mathbf{B}+\nabla p = \mathbf{f}
\end{equation}
with:
\begin{equation}
\label{eq:b}
\mathbf{B}
=\rho\mathbf{A}_t+\mu\,\mathbf{\omega}
\quad.
\end{equation}

\begin{figure}[t]
\begin{center}
\includegraphics{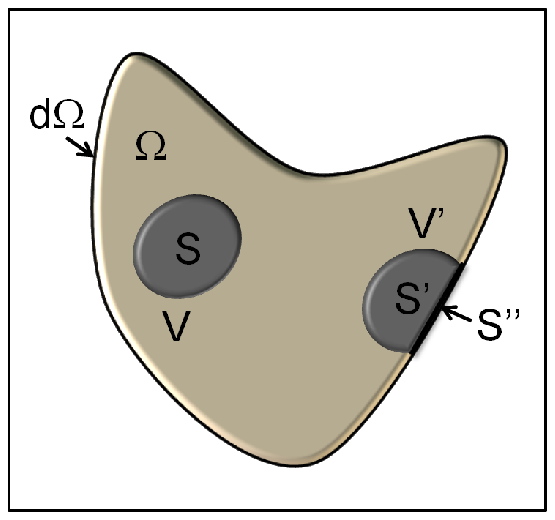}
\caption{Scheme} \label{fig:integrals}
\end{center}
\end{figure}

The question now is: what form of boundary conditions can be used to compute $p$ as the scalar quantity entering the Helmholtz decomposition~(\ref{eq:h2})? To answer it, let us first examine how this pressure problem can be given an integral representation. We thus consider a volume $V\in\Omega$, with boundary $S$, that does not intersect the domain boundary $\partial\Omega$. Denoting $\mathbf{n}$ the normal vector to any surface element $\d S$, we have:
\begin{equation}
\int_S \mathbf{n}\cdot\nabla p \,\,\d S=\int_V\div \mathbf{f}\,\,\d V
\end{equation}
This of course is an integral form of the Poisson problem $\Delta p=\div\mathbf{f}$.
Second, we consider a volume $V'$ that does intersect the domain boundary. As sketched on Fig.~\ref{fig:integrals}, the boundary of $V'\cap\Omega$ can be decomposed into two components: a surface $S'$ which lies in the interior of $\Omega$ and a surface $S''\in\partial\Omega$. We then find:
\begin{equation}
\int_{S'} \mathbf{n}\cdot\nabla p \,\,\d S=\int_{V'}\div \mathbf{f}\,\,\d V -\int_{S''}\,\mathbf{n}\cdot\left(\mathbf{f}-\rot\mathbf{B}\right)\,\d S
\end{equation}

These integrals representation corresponds to respectively equations~(\ref{integral:p:2d:closed}) and~(\ref{integral:p:2d}) in the 2D case: the problem is that here, they involve the boundary values of a vector field $\mathbf{B}$ -- versus a scalar field in 2D -- which depends on the gauge prescription: we a priori do not know how to compute $\mathbf{B}$ and need to guarantee that any gauge fixing will lead to an identical discretization for the pressure problem.

We first write:
$$
\int_{S''}\,\mathbf{n}\cdot\rot\mathbf{B}\,\d S=\int_{\Gamma''} \mathbf{B}\cdot\d \mathbf{l}
$$
with $\Gamma''$ the curve enclosing $S''$, and note that the integral on the rhs involves only the tangential components of $\mathbf{B}$.
To study the influence of gauge fixing, we consider one $\mathbf{A}_s$ such that $\mathbf{u}=\rot\mathbf{A}_s$: all valid vector potentials for $\mathbf{u}$ are of the form  $\mathbf{A}=\mathbf{A}_s+\nabla\varphi$. We then use the Coulomb gauge, $\div\mathbf{A}=0$, which amounts to requiring that $\varphi$ verifies $\Delta\varphi=-\nabla\cdot\mathbf{A}_s$. This Poisson equation can now be provided boundary conditions. The choice of any Dirichlet condition of $\varphi$  -- up to an irrelevant constant -- amounts to (i) fixing the components of $\nabla\varphi$ tangent to the interface plus (ii) the additional constraint that for any closed curve $\Gamma\in\partial\Omega$: $\int_\Gamma\nabla\varphi\cdot\mathbf{dl}=0$, with $\mathbf{dl}$ the normalized line element along $\Gamma$. The second condition guarantees that we can define $\varphi$ via integral equations of the form $\varphi_B-\varphi_A=\int_A^B\nabla\varphi\cdot\mathbf{dl}$ where integrals are taken along curves $\in\partial\Omega$ running from $A$ to $B$. The Dirichlet condition on $\varphi$ is thus equivalent to the prescription of the tangential components of $\mathbf{A}$, provided for any closed contour $\Gamma\in\partial\Omega$ they verify:
\begin{equation}
\int_\Gamma\mathbf{A}\cdot\mathbf{dl}=
\int_\Gamma\mathbf{A}_s\cdot\mathbf{dl}=\int \mathbf{u}\cdot\mathbf{n}\d S
\label{eq:mass:interface}
\end{equation}
In short, any choice of the tangential components of $\mathbf{A}$ can be made, provided they are consistent with equation~(\ref{eq:mass:interface}), i.e. they enforce the correct mass fluxes through the interface. In practice, near rigid interfaces, it will always be possible to compute analytically such components of a valid vector potential, without the need to compute $\mathbf{A}$ entirely. To fix $\mathbf{B}\cdot\d\mathbf{l}$ on the domain boundary, it remains to evaluate the tangential components of the vorticity there, which like in the 2D case can be performed using derivatives of the components of the velocity field.

We thus find that the treatment of 3D flows should follow exactly along similar lines to those presented here in 2D.

\section{Conclusion}

We have here shown that by viewing the Stokes equation as a Helmholtz decomposition of the field of bulk forces, it was possible to construct a discretization based on the physical $p$ and $\mathbf{u}$ fields, which accurately enforces mass conservation. The resulting evaluation of the pressure field in then devoid of numerical errors due to spurious mass sources and sinks and has been here shown to converge with second order even near boundaries. Our implementation relies on the introduction of ghost points, yet seems to correctly capture singularities of the pressure e.g. near obstacles such as in a Darcy flow.

The present work does not pretend to be complete, some limitations can be immediately identified, which may be more or less challenging:
\begin{enumerate}
\item we have here focussed on the Stokes flow: this is not a strong problem as implementing the non-linear terms of the Navier-Stokes equation does not introduce significant new technicalities;
\item we have relied on time-explicit discretizations so as to focus on the difficulties arising from the evaluation of the pressure field: extending our method to time-implicit scheme might require significant revisions in the treatment of the vorticity field, however, and this we must leave for future works;
\item finally, we have focussed our discussion on the case of 2D problems, but have also argued the extension to the 3D case should be rather immediate and mostly technical.
\end{enumerate}
We hope, however, that our work can open new routes to considering the difficult question of implementing direct numerical simulations of fluid flows in the presence of interfaces.

\section*{Acknowledgements}
We acknowledge many discussions with Mikha\"el Balabane, Jean-Christophe Nave, and Ruben R. Rosales.


%
%

\end{document}